\documentclass[traditabstract]{aa}
\usepackage{txfonts}
\usepackage{graphicx}
\usepackage{natbib}
\newcommand{\AK}[1]{}
\usepackage{array}
\bibpunct{(}{)}{;
}{a}{}{,}

\begin{document}

\title{Young open cluster IC 4996 and its vicinity: multicolor
photometry and \textit{\textbf{Gaia}} DR2 astrometry\thanks{Tables 1, 2,
4 and 5 are only available in electronic form
at the CDS via anonymous ftp to cdsarc.u-strasbg.fr (130.79.128.5)
or via http://cdsweb.u-strasbg.fr/cgi-bin/qcat?J/A+A/}}

\author
{V. Strai\v{z}ys
\inst{1}$^{}$\thanks{e-mail:vytautas.straizys@tfai.vu.lt}
\and R. P. Boyle \inst{2}
\and K. Mila\v{s}ius \inst{1}
\and K.~\v{C}ernis \inst{1}
\and M. Macijauskas \inst{1}
\and U. Munari \inst{3}
\and R. Janusz \inst{4}
\and J. Zdanavi\v{c}ius \inst{1}
\and \\K. Zdanavi\v{c}ius \inst{1}
\and M. Maskoli\={u}nas \inst{1}
\and S. Raudeli\={u}nas \inst{1}
\and A. Kazlauskas \inst{1}
}
\institute{Institute of Theoretical Physics and Astronomy, Vilnius
University, Saul\.etekio al. 3, Vilnius LT-10257, Lithuania
\and Vatican Observatory Research Group, Steward Observatory,
 Tucson, AZ 85721, U.S.A.
\and INAF Astronomical Observatory of Padova, I-36012, Asiago (VI),
Italy
\and Vatican Observatory, V-00120, Vatican City State}

\date{Received 2018 ...; accepted 2019-01-13; in original form
2018 ......}

\label{firstpage}

\abstract{The open cluster IC 4996 in Cygnus and its vicinity are
investigated by applying a two-dimensional photometric classification of
stars measured in the Vilnius seven-color photometric system.  Cluster
members are identified by applying distances based on the {\it Gaia} DR2
parallaxes and the point vector diagram of the {\it Gaia} DR2 proper
motions.  For some B-type stars, spectroscopic MK types are also
obtained from the Asiago spectra and collected from the literature.  New
parameters of the cluster are derived.  The interstellar extinction
$A_V$ covers a wide range of values, from 1.3 to 2.4 mag; the mean value
in the central part of the cluster is 1.8 mag.  The cluster distance is
1915\,$\pm$\,110 pc, and its age is within 8--10 Myr. The cluster
exhibits a long sequence from early-B to G stars, where stars cooler
than B8 are in the pre-main-sequence stage.  The plot of extinction
versus distance shows a steep rise of $A_V$ up to 1.6 mag at 700--800
pc, which is probably related to dust clouds at the edge of the Great
Cygnus Rift.  The next increase in extinction by an additional 0.8 mag
at $d$\,$\geq$\,1.7 kpc is probably related to the associations Cyg OB1
and Cyg OB3.  The cluster IC 4996 does not belong to the Cyg OB1
association, which is located closer to the Sun, at 1682\,$\pm$\,116 pc.
It seems likely that the cluster and the surrounding O-B stars have a
common origin with the nearby association Cyg OB3 since {\it Gaia} data
show that these stellar groups are located at a similar distance.}


\titlerunning {Young open cluster IC 4996}
\authorrunning{V. Strai\v{z}ys et al.}
\maketitle

\section{Introduction}

The new astrometric data obtained by the {\it Gaia} mission
\citep{Gaia2016} enable verifying the reality of large systems of
massive stars that are known as OB associations.  Many of them contain
young open clusters, and their relation to the surrounding associations
is an important clue for understanding their common evolution.  One such
system is the Cyg OB1 association, which is considered to contain the
young clusters M29 (NGC 6913), IC 4996, Berkeley 86, and Berkeley 87;
see the recent review by \citet{Reipurth2008}.  According to
\citet{Humphreys1978, Humphreys1984, Blaha1989, Garmany1992} and
\citet{Melnik2017}, the Cyg OB1 association contains about 70--75 OB
stars and M supergiants of magnitudes 7--11, scattered in the
4.5$\degr$\,$\times$\,4.5$\degr$ area.

However, according to the analysis of \citet{Cantat2018} (hereafter
CG18) of the open cluster population based on {\it Gaia} DR2 astrometry,
the distances to the four clusters located within the Cyg OB1 area do
not coincide, even when their error bars are taken into account.  While
the distances to M29, Berkeley 86, and Berkeley 87 are sufficiently
close (1719 pc, 1703 pc, and 1661 pc), the distance to IC 4996 is 1937
pc, and it exceeds the average distance to the three other clusters by
243 pc, which is much larger than the distance uncertainties (50--100
pc).  Thus, a new analysis of the members of cluster IC 4996, based on
their spectral and photometric classifications and also on their
location in the Hertzsprung-Russell (HR) diagram is important.

In the direction of IC 4996, the Milky Way shows a strong and variable
interstellar reddening.  This cluster is located only a few degrees from
the Great Cygnus Rift, the concentration of dust clouds at a distance of
700--800 pc.  Photometric studies of the cluster and its vicinity were
conducted in the following systems:  {\it UBV} \citep{Hoag1961,
Delgado1998}, {\it BV} \citep{Maciejewski2007}, {\it BVRI}
\citep{Vansevicius1996}, {\it RGU} \citep{Purgathofer1961, Becker1963},
Geneva \citep{Nicolet1981}, Vilnius \citep{Sudzius1976, Pucinskas1982,
Vansevicius1989}, {\it uvby}H$\beta$ \citep{Alfaro1985}, and {\it JHK}
\citep{Bhavya2007, Kharchenko2013, Buckner2016}.  Despite numerous
photometric and spectral investigations of the cluster, its distance and
age gave contradicting results.  For example, its distance in various
sources covers the range from 1.67 kpc to 2.40 kpc, and the ages of the
cluster are given from 6 Myr to 9 Myr. These differences may be related
to differing sets of accepted cluster members, uncertainties in spectral
and photometric classifications of stars, variable interstellar
reddening, unresolved duplicity of cluster stars and their
contamination by field stars.  Although \citet{Delgado1998,
Delgado1999}, \citet{Zwintz2006}, and \citet{Bhavya2007} identified a
few stars that were suspected to be pre-main-sequence objects of
spectral classes A and F, the presence of the whole sequence of A-F-G
stars remains unconfirmed.

To classify individual stars in temperatures and gravities (or into
spectral and luminosity classes), either spectroscopy or photometry in
the Vilnius seven-color system with mean wavelengths at 345 ($U$), 374
($P$), 405 ($X$), 466 ($Y$), 516 ($Z$), 544 ($V$), and 656 ($S$) nm can
be applied \citep{Straizys1992}.  This system gives a two-dimensional
classification of stars of all spectral types (from O to M) in the
presence of variable interstellar reddening.  In some temperature
ranges, the system can also give indications of metallicity, binarity,
or peculiarity.

In this article, we attempt to determine the parameters of IC 4996 with
new spectroscopic MK types of the brightest stars and two-dimensional
photometric spectral types of fainter stars based on their CCD
photometry in the Vilnius system and individual dereddening.  To
identify cluster members in the 13$\arcmin$\,$\times$\,13$\arcmin$ area,
we apply their data from the {\it Gaia} Data Release 2 (hereafter {\it
Gaia} DR2):  proper motions from \citet{Gaia2018} and distances from
\citet{Bailer2018}.  To determine the cluster membership, we used the
Gaussian mixture model (GMM)
program\footnote[1]{https://github.com/jobovy/extreme-deconvolution} by
\citet{Bovy2009}.  We also investigate interstellar extinction in the
foreground and background of the cluster.  The age of the cluster is
estimated using the effective temperatures and bolometric luminosities
determined from photometry and spectral types, and a set of the Padova
isochrones.  We also estimate the relation between the cluster IC 4996
and the associations Cyg OB1 and Cyg OB3.

\section{Photometric data and spectral types}

The investigated 13$\arcmin$\,$\times$\,13$\arcmin$ area (Figure 1) is
centered on the cluster IC 4996 at RA (J2000) = 20:16:30, DEC (J2000) =
+37:38.  CCD exposures with the filters of the Vilnius system were
obtained in 2011--2013 with the 1.8 m Vatican Advanced Technology
Telescope (VATT) on Mt. Graham, Arizona, using the STA0500A CCD camera
with a 4k\,$\times$\,4k chip containing pixel sizes of 15\,$\times$\,15
$\mu$m.  In each filter, about 20 exposures of different duration were
obtained.  The CCD frames were processed with the IRAF program package
in the aperture mode.  First, all measurements in each filter were
transformed into one instrumental system, taking into account the
different air mass of every exposure and averaging the data.  Next, the
instrumental magnitudes $m_V$ and color indices $C_{U-V}$, $C_{P-V}$,
$C_{X-V}$, $C_{Y-V}$, $C_{Z-V}$, and $C_{V-S}$ were formed and
transformed into $V$, $U-V$, $P-V$, $X-V$, $Y-V$, $Z-V$, and $V-S$ of
the standard system.

Equations for the transformation of $V$ magnitudes and six color indices
from the instrumental to the standard system were obtained using
photoelectric observations of 20 stars from \citet{Vansevicius1989} and
12 stars observed by one of the authors (A.K., unpublished) at the
Maidanak Observatory in 1978, all located in the CCD field.  The
magnitudes $V$ and color indices of seven stars in common to both lists
were averaged, and a set of 24 standards was composed (after excluding
the variable V1922 Cyg).  Because the response curves of the
instrumental and the standard systems are quite similar, the linear
color equations were sufficient.  Only the zero-points and small color
terms were considered.  No deviations depending on the luminosity and
interstellar reddening were detected.  Transformation errors are about
$\pm$\,0.02 mag.

Table 1 contains the results of photoelectric photometry of 76 stars
located within $\sim$\,25$\arcmin$ from the cluster center.  The
following data are given:  star number, equatorial coordinates J2000,
magnitude $V$, color indices $U$--$V$, $P$--$V$, $X$--$V$, $Y$--$V$,
$Z$--$V$, and $V$--$S$, photometric spectral types in the MK system and
the indication of duplicity within $\sim$\,6$\arcsec$.  The numbers of
stars in Table 1 are marked with the initials AK to avoid
misidentifications with Table 2. Photometric spectral types are given in
lower-case letters to distinguish them from the spectroscopic classes.


\begin{figure}
\resizebox{\hsize}{!}{\includegraphics{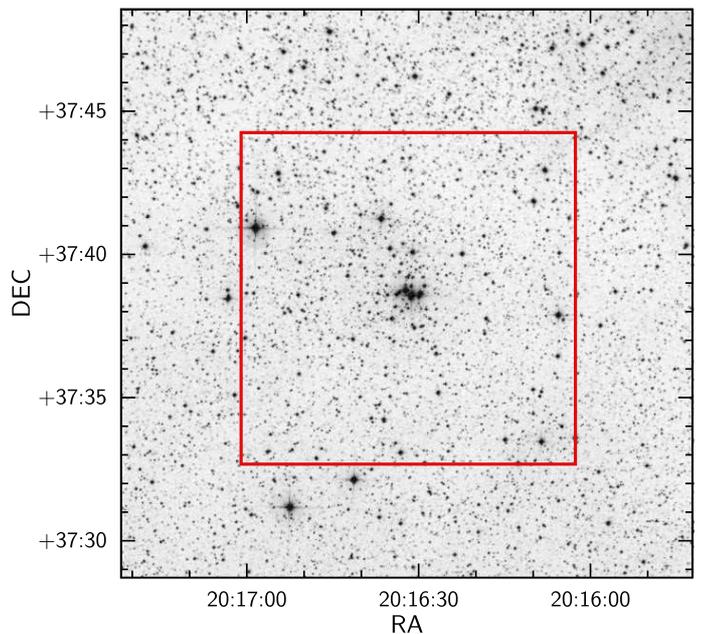}}
\vskip1mm
\caption{ 20$\arcmin$\,$\times$\,20$\arcmin$ area covered by the
stars for which the new spectroscopic and photometric MK
classifications were obtained. The red square shows the CCD
area. The background is from DSS RED2.}
\end{figure}


\begin{table*}
\caption{First five lines of the catalog of 76 stars in the IC 4996
area containing the results of photoelectric photometry with the 48 cm
telescope of the Maidanak Observatory and their photometric spectral
types.  The full table is available at the CDS.}
\tabcolsep=2pt
\label{table1}
\centering
\begin{tabular}{rccrcccccclc}
\hline\hline
\noalign{\vskip0.5mm}
 No. &  RA\,(J2000) & DEC\,(J2000) & $V$~~~~ &  $U$--$V$ &
$P$--$V$ & $X$--$V$ & $Y$--$V$ & $Z$--$V$ & $V$--$S$ & Photom. & Bin. \\
 & h~~~~m~~~~~s~ & $\circ$~~~~$\prime$~~~~$\prime\prime$
 & mag~ & mag & mag & mag & mag  & mag & mag & sp. type & \\
\hline
\noalign{\vskip 0.5mm}
 AK1 &  20:15:07.46 & +37:38:26.7 &  10.025 & 1.846 & 1.298 & 0.522 & 0.199 & 0.067 & 0.131 & a0.5 IV  &  \\
 AK2 &  20:15:09.02 & +37:33:05.2 &  11.508 & 2.242 & 1.714 & 1.140 & 0.467 & 0.174 & 0.470 & f6 IV    &  \\
 AK3 &  20:15:17.57 & +37:31:43.9 &  12.167 & 2.333 & 1.823 & 1.259 & 0.547 & 0.238 & 0.490 & f6 V     &  \\
 AK4 &  20:15:21.34 & +37:35:53.1 &   9.583 & 1.856 & 1.260 & 0.500 & 0.195 & 0.062 & 0.135 & b8.5 V   &  \\
 AK5 &  20:15:23.33 & +37:33:30.7 &   9.947 & 2.232 & 1.509 & 0.646 & 0.260 & 0.089 & 0.217 & a1.5 III &  \\
\hline
\end{tabular}
\end{table*}


\begin{table*}
\caption{Five lines of the catalog of 1337 stars in the IC 4996 area
that contain the results of CCD photometry with the VATT and
photometric spectral types. The full table is available at the CDS.}
\tabcolsep=2pt
\label{table2}
\centering
\begin{tabular}{rccrcccccclc}
\hline\hline
\noalign{\vskip0.5mm}
 No. &  RA\,(J2000) & DEC\,(J2000) & $V$~~~~ & $U$--$V$ &
$P$--$V$ & $X$--$V$ & $Y$--$V$ & $Z$--$V$ & $V$--$S$ & Photom. & Bin. \\
 & h~~~~m~~~~~s~ & $\circ$~~~~$\prime$~~~~$\prime\prime$
 & mag~ & mag & mag & mag & mag & mag & mag & sp. type &  \\
\hline
\noalign{\vskip 0.5mm}
 601 &  20:16:31.21  & +37:40:02.3  & 10.498 &  1.573 &  1.210 &  0.902 &  0.488 &  0.190 &  0.423 &  b1   III-IV &      \\
 602 &  20:16:31.30  & +37:37:29.7  & 14.100 &  2.894 &  2.147 &  1.257 &  0.656 &  0.250 &  0.518 &  b8   V      & **   \\
 603 &  20:16:31.33  & +37:33:08.5  & 16.974 &  3.471 &  2.846 &  1.988 &  0.951 &  0.332 &  0.875 &  f8   III    &      \\
 604 &  20:16:31.38  & +37:44:37.7  & 17.591 &  3.689 &  2.944 &  1.931 &  0.962 &  0.360 &        &  f0   V      & **   \\
 605 &  20:16:31.42  & +37:38:51.1  & 12.694 &  2.237 &  1.718 &  1.078 &  0.561 &  0.213 &  0.483 &  b5   IV-V   &      \\
\hline
\end{tabular}
\end{table*}

The catalog with the results of CCD photometry for 1337 stars down to
$V$\,$\approx$\,18.8 mag and the results of photometric classification
for about 70\% of them are given in Table 2, which is available online
at the CDS.  The uncertainties of magnitudes $V$ and all color indices,
which take into account the measurement errors and the errors of
transformation to the standard system, are $\leq$\,0.03 mag down to $V$
= 18 mag.  If the uncertainty is larger than 0.05 mag, color index is
not given.  The visual binaries with asymmetric images or components
within 2.5$\arcsec$ are marked with two asterisks in the last column.
No photometric classification is given for them.

For the two-dimensional classification of stars, two codes, COMPAR and
NORMA, were applied.  The COMPAR code, described in our previous
publications (e.g., \citet{Straizys2013}), matches 14 different
interstellar reddening-free $Q$-parameters of a program star to those of
about 8000 standard stars of various spectral and luminosity classes,
metallicities, and peculiarities.  The NORMA code has recently been
developed by one of the authors (K.Z.).  It uses a set of 808 standards
formed from the intrinsic color indices of the Vilnius system for
different spectral and luminosity classes.  More details are given in
\citet{Straizys2018}.

The classification accuracy has recently been described by
\citet{Straizys2018}.  Briefly, the accuracy of spectral class is on the
order of 1--2 decimal subclasses.  The accuracy of the luminosities for
B8-A-F-G5 stars of luminosity classes V-IV-III is about one luminosity
class.  For K-type stars, the accuracy is about 0.5 of a spectral subclass
and 0.5 of a luminosity class.  To classify K- and M-type
stars, the ultraviolet $U$--$V$ and $P$--$V$ color indices are not
essential because the photometric temperature and luminosity criteria are
sufficiently strong in the passbands from $X$ to $S$.  Because the
classification accuracy for the stars of $V > 18$ mag is somewhat lower,
no luminosity classes are given for them in most cases (except for
K-type stars).

To verify the results of photometric classification, 20 stars of
spectral classes B and A, brighter than $V$ = 12.1 mag, were observed
spectroscopically with the 1.22 m telescope at the Asiago Observatory.
These stars are located up to 18$\arcmin$ from the cluster center.  The
spectra cover a wavelength range of 350--680 nm and have a dispersion of
2.31 \r{A} in pixel.  The stars were classified by fitting them to the
library of MK standards taken with the same instrumental configuration,
as well as with the MKCLASS computer program by \citet{Gray2014}.  Table
3 gives their MK types determined from the Asiago spectra, photometric
spectral types, and the types collected from the literature and given in
the SIMBAD database.  The agreement between spectroscopic and
photometric spectral classes is within 1--2 decimal subclasses and one
luminosity class.  Stars 174, 573, 661, and 1140 are Washington Double
Star Catalog (WDS) visual binaries \citep{Worley1997}, so their
classifications by these two methods are less reliable.


\begin{table}
\caption{Stars classified from the Asiago spectra.  The first column
gives the star number from Tables 1 or 2. Asterisks at the numbers
designate binary stars.  The stars are listed in the order of right
ascension.  For comparison, we list photometric spectral types and
spectral types collected at the SIMBAD database.}
\label{table3}
\begin{tabular}{crlll}
\hline\hline
\noalign{\vskip 0.5mm}
No. & $V$~~        & Sp         & Sp            & Sp         \\
    & mag          & Asiago     & photom.       & other      \\
\hline                                                                                    
\noalign{\vskip 0.5mm}
 AK1      & 10.025 & B9 V       &  a0 V         &    B9 V             \\
 AK5      &  9.947 & A2 IV      &  a2 IV-V      &    B9.5 V           \\
  46      & 10.887 & A5 IV      &  a5 V         &    --               \\
 163      & 10.185 & B1 IV      &  b1           &    OBe, B2 II:      \\
 174*     & 11.310 & B3 III     &  b3 IV-V:     &    OB               \\
 553      & 12.015 & B1 IVe     &  b3:          &    B3               \\
 AK36*    &  9.268 & B1 III     &  b1 IV        &    B0.5 V:          \\
 573*     &  9.765 & B1 III     &  b1 III       &    B1 III, B1 IV    \\
 AK37     & 10.005 & B1 IV      &  b2 IV        &    B0.5 V           \\
 601      & 10.498 & B1 IV      &  b1 IV        &    B3               \\
HD 193007* &  7.974 & B0 II     &  --   &  B0.5\,II, B0\,III          \\
BD+37 3862 & 8.982 & B1 IIIe    &  --   &    B0\,V, B0\,IVp           \\
 661*     & 11.800 & B3 III     &  b3 IV:       &    --               \\
 668      & 11.341 & B3 III     &  b2 IV        &    --               \\
 735      &  9.488 & B1 Ib      &  b1 III       &    OB               \\
 AK46     &  7.776 & A9 III     &  a9 IV        &    A9 III           \\
 AK47     &  8.756 & A3 IV      &  a3 IV        &    A3               \\
 910      & 10.863 & B8 III     &  b8 IV        &    B5               \\
1140*     & 11.388 & B7 II      &  b8           &    B5               \\
1266      & 12.086 & A0 III     &  a1 IV        &    --               \\
\hline
\end{tabular}
\end{table}

\section{Interstellar extinctions and distances}

In the COMPAR code, color excesses of stars are calculated by the
equation
\begin{equation}
E_{Y-V} = (Y-V)_{\rm obs} - (Y-V)_0,
\end{equation}
where $(Y-V)_0$ are the intrinsic colors taken from
\citet{Straizys1992}.  In the NORMA code, color excesses of stars for
all six color indices, given in Tables 1 and 2, are calculated:
\begin{equation}
E_{m-V} = (m-V)_{\rm obs} - (m-V)_0,
\end{equation}
where $(m-V)_{\rm obs}$ and $(m-V)_0$ are the observed and intrinsic
color indices, respectively.  Then $E_{U-V}$, $E_{P-V}$, $E_{X-V}$,
$E_{Z-V}$ and $E_{V-S}$ are transformed into $E_{Y-V}$ and all the six
values are averaged.  If the ultraviolet color indices $U$--$V$ and
$P$--$V$ are not available, then only four values of color excess are
averaged. The $E_{Y-V}$ values are transformed into the extinctions
$A_V$ by the equation
\begin{equation}
A_V = 3.86\,E_{Y-V},
\end{equation}
where the coefficient 3.86 was determined from the ratios
$E_{V-J}/E_{B-V}$, $E_{V-H}/E_{B-V}$, and $E_{V-K_s}/E_{B-V}$ for B-
and A-type stars, as described in \citet{Straizys2014}.  In the IC 4996
area, this is very close to the coefficient obtained for the direction
to the cluster M29. Both are slightly lower than the normal value, 4.16,
which corresponds to 3.15 in the {\it B,V} system.  This normal value of
the coefficient would give values of $A_V$ that are higher by a factor
of 1.08.  The typical uncertainty of $A_V$ is $\sim$\,0.10 mag because
of the observational errors of $Y$--$V$ and the errors of the intrinsic
$(Y-V)_0$ colors.  The extinction error mostly depends on the errors of
spectral classes, while the error of the luminosity class is much less
important.

Distances to the stars were taken from the catalog of
\citet{Bailer2018}, where they have been calculated from the {\it Gaia}
DR2 parallaxes inverted to distances, taking the nonlinearity of the
transformation and the asymmetry of the resulting probability
distribution into account.


\begin{figure}
\resizebox{\hsize}{!}{\includegraphics{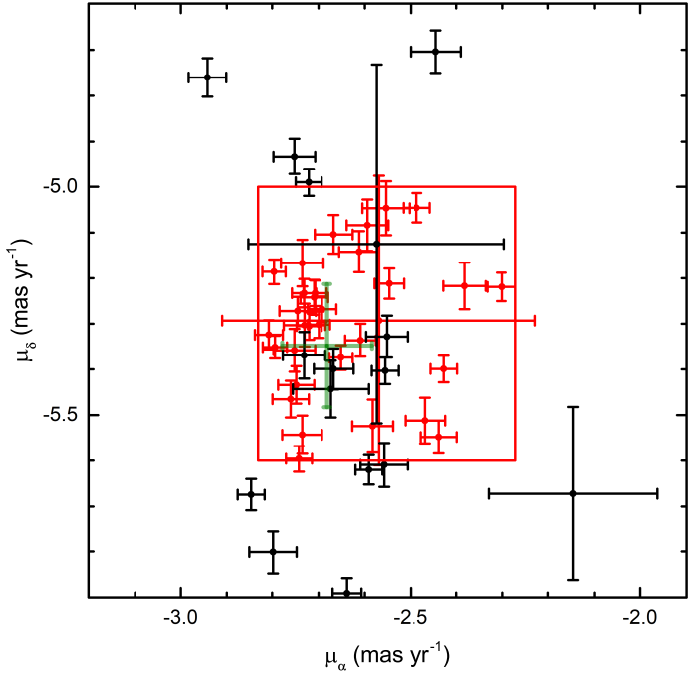}}
\caption{Vector point diagram of proper motions for B-type stars down to
$V$ = 14 mag within 13$\arcmin$\,$\times$\,13$\arcmin$ around IC 4996
from {\it Gaia} DR2.  The red rectangle denotes the accepted
location of the cluster members, which are shown as red dots and error
crosses.  The green cross designates the center of the cluster members
in the 7$\arcmin$\,$\times$\,7$\arcmin$ area according to CG18.}
\vskip1mm
\end{figure}

\section{Cluster membership, distance, and age}

To select possible members of IC 4996, we applied {\it Gaia} DR2 proper
motion components from \citet{Gaia2018} and distances from
\citet{Bailer2018}.  From Tables 1 and 2, we first selected all B-type
stars down to $V$ = 16 mag that are located in the
13$\arcmin$\,$\times$\,13$\arcmin$ area centered on the cluster.  These
stars were plotted in the vector point diagram (VPD) $\mu_{\alpha}$
versus $\mu_{\delta}$; here $\mu_{\alpha}$ contains a factor of
cos\,$\delta$.  Four bright stars, HD 228657, HD 193007, HD 193076, and
BD+37 3862, measured earlier in the Vilnius system by
\citet{Vansevicius1989}, were also added.  These 64 stars in the VPD
cover quite a large area from +2 to --5 mas\,yr$^{-1}$ in $\mu_{\alpha}$
and from --2 to --8 mas\,yr$^{-1}$ in $\mu_{\delta}$.  About 50\% of
these stars are concentrated in the rectangle limited by the coordinates
from --2.3 to --2.8 mas\,yr$^{-1}$ in $\mu_{\alpha}$ and from --5.0 to
--5.6 mas\,yr$^{-1}$ in $\mu_{\delta}$, which is shown in red in Figure
2. The histogram of distances to the stars that are located in the red
rectangle shows that 32 of them are concentrated in the 1.70--2.15 kpc
distance range,  with the maximum number of stars at a distance of 1.9
kpc.
We accept that these 32 B-type stars are the cluster members; their
distance range covers about $\pm$\,2$\sigma$ of the distance errors.  In
Figure 2 these stars are shown as red dots with their error crosses.
The black dots are field B-stars of the foreground and background.  More
field stars are located outside the red rectangle and outside the
diagram.

The red rectangle in Figure 2, defined by B-type stars, was used to
select cluster members with lower temperature.  By applying the {\it
Gaia} distances and proper motions, we identified within the
13$\arcmin$\,$\times$\,13$\arcmin$ area 40 stars from A0 to K2 as
possible cluster members (Table 4).  To verify the reliability of our
selection, we used the GMM program from \citet{Bovy2009}.  We performed
clustering in the 13$\arcmin$\,$\times$\,13$\arcmin$ area using three
parameters (parallax and proper motions) for each star.  The GMM can
select a local instead of a global minimum, therefore at least 100 runs
were made using random initial values of the GMM parameters.  As the
cluster membership probability $p$, we used the hyperparameter $q_{ij}$
introduced by \citet{Bovy2009} (their Eq.  (16)).  This
probability is given in the last column of Table 4. The GMM procedure
allowed us to specify the center of cluster members in the VPD:
$\mu_{\alpha}$ = --2.643\,$\pm$\,0.145 mas\,yr$^{-1}$ and $\mu_{\delta}$
= --5.311\,$\pm$\,0.171 mas\,yr$^{-1}$.  This center is quite close to
the center of the red rectangle in Figure 2. All cluster members in the
rectangle have membership probabilities $p$\,$\geq$\,0.7.  In the list
of 1337 stars from Table 2, the GMM program also identified 53
additional stars with a membership probability $p$\,$\geq$\,0.7, but
these stars are either fainter than $V$ = 16 mag or do not have reliable
spectral classifications.  The final list of selected 72 cluster members
is given in Table 4. This list contains 8 stars that were listed as
possible pre-main-sequence objects by \citet{Delgado1999}.

Using the {\it Gaia} DR2 proper motions and parallaxes, CG18 estimated
membership probabilities and distances for 1229 clusters, including IC
4996.  In this cluster, they identified 79 stars within a
7$\arcmin$\,$\times$\,7$\arcmin$ area down to $G$ = 18 mag with
membership probabilities $>$\,0.5.  The distance to the cluster based on
the {\it Gaia} parallaxes was found to be 1937 pc.  The following
average proper motion value of the cluster was found:  $\mu_{\alpha}$ =
--2.681\,$\pm$\,0.097 mas\,yr$^{-1}$ and $\mu_{\delta}$ =
--5.348\,$\pm$\,0.135 mas\,yr$^{-1}$.  In Figure 2 it is shown as a
green dot and the error cross.  This result was based on all member
stars, regardless of their spectral classes or temperatures.  Comparing
our list with the list of 79 possible members of CG18, we found 50 stars
in common with membership probabilities $>$\,0.5.  Other members of CG18
are too faint to be included in Table 4 or are binaries and do not have
reliable spectral types.  In our list of 72 possible members, 5 stars in
CG18 have lower membership probabilities, and 17 stars lie outside the
7$\arcmin$\,$\times$\,7$\arcmin$ area of CG18.


\begin{table*}
\caption{First five stars of the list of IC 4996 members with
their classification and intrinsic parameters. The last six columns
give the proper motions from {\it Gaia} DR2, the estimated distances
from \citet{Bailer2018}, and the membership probability from GMM. The
full table is available at the CDS.}
\label{table4}
\centering
\tabcolsep=2.5pt
\begin{tabular}{rcccccccccccccccc}
\hline\hline
\noalign{\vskip0.5mm}
No. & RA(J2000) & DEC(J2000) & $V$~~ & Sp. type  & $A_V$ & $(Y-V)_0$ &
$V_0$ & $BC$ & log\,$T_{\rm eff}$ & log\,$L/L_{\odot}$ & $\mu_{\alpha}$ &
$\sigma(\mu_{\alpha})$ & $\mu_{\delta}$ & $\sigma(\mu_{\alpha})$
& $d$ (pc) & $p$ \\
\hline
\noalign{\vskip0.5mm}
   34  &   20:16:02.16 & +37:41:30.1 & 15.929  &  a7 V   &  2.34  &  0.27  &  13.59  & +0.02   & 3.903  & 1.004 & -2.713 & 0.057 & -5.625 & 0.066 & 1811  & 0.90 \\
  136  &   20:16:07.40 & +37:42:34.6 & 14.442  &  a0 V   &  1.84  &  0.17  &  12.60  & --0.25  & 3.982  & 1.508 & -2.472 & 0.032 & -5.226 & 0.033 & 1919  & 0.97 \\
  162  &   20:16:08.55 & +37:35:08.2 & 13.662  &  b8 V   &  2.20  &  0.11  &  11.46  & --0.80  & 4.107  & 2.184 & -2.796 & 0.025 & -5.186 & 0.027 & 1843  & 0.96 \\
  203  &   20:16:10.93 & +37:34:53.1 & 14.948  &  a0 V   &  2.27  &  0.18  &  12.68  & --0.20  & 3.978  & 1.456 & -2.540 & 0.036 & -5.469 & 0.037 & 2042  & 0.96 \\
  289  &   20:16:15.44 & +37:40:15.4 & 15.135  &  a3 V   &  1.78  &  0.22  &  13.36  & --0.03  & 3.949  & 1.116 & -2.750 & 0.044 & -5.411 & 0.046 & 2058  & 0.96 \\
\hline
\end{tabular}
\end{table*}


\begin{table*}
\caption{First five stars of the list of the association Cyg OB1
members with their proper motions from {\it Gaia} DR2 and the distances
$d$ from \citet{Bailer2018}.  Magnitudes $V$ and spectral types are from
\citet{Humphreys1984}.  The full table is available at the CDS.}
\label{table5}
\centering
\tabcolsep=2.5pt
\begin{tabular}{cccrlccccc}
\hline\hline
\noalign{\vskip0.5mm}
No. & RA(J2000) & DEC(J2000) & $V$~~ & Sp. type  &  $\mu_{\alpha}$ &
$\sigma(\mu_{\alpha})$ & $\mu_{\delta}$ & $\sigma(\mu_{\alpha})$
& $d$ (pc) \\
\hline
\noalign{\vskip0.5mm}
  HD 192079  & 20:11:44.99 & +37:32:59.9 & 8.77 & B2 III & -2.751 & 0.055 & -5.706 & 0.055 & 1672.2  \\
  HD 228456  & 20:14:02.31 & +36:48:07.0 & 9.86 & B2 IV  & -2.728 & 0.078 & -5.680 & 0.087 & 1693.9  \\
  HD 228461  & 20:14:06.42 & +38:14:38.3 & 9.47 & B2 II  & -2.563 & 0.040 & -5.126 & 0.042 & 1691.2  \\
  HD 228533  & 20:14:48.75 & +38:27:14.6 & 9.08 & B0 III & -2.718 & 0.053 & -4.992 & 0.048 & 1792.4  \\
  HD 228543  & 20:14:56.62 & +38:08:18.3 & 8.76 & B2 II  & -3.520 & 0.053 & -5.572 & 0.059 & 1646.3  \\
\hline
\end{tabular}
\end{table*}


\begin{figure}
\resizebox{\hsize}{!}{\includegraphics{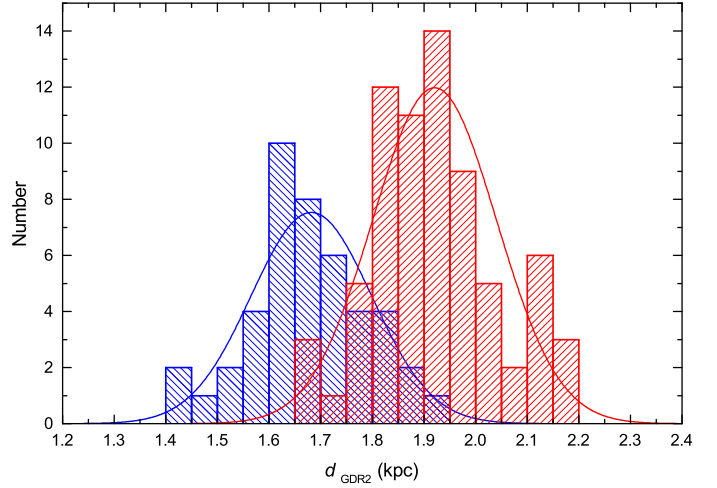}}
\caption{Distance histograms for 72 probable
members of the cluster IC 4996 (red) and 44 probable members of the
association Cyg OB1 (blue). The Gaussian curves are shown for both
groups of stars.}
\vskip1mm
\end{figure}


\begin{figure}
\resizebox{\hsize}{!}{\includegraphics{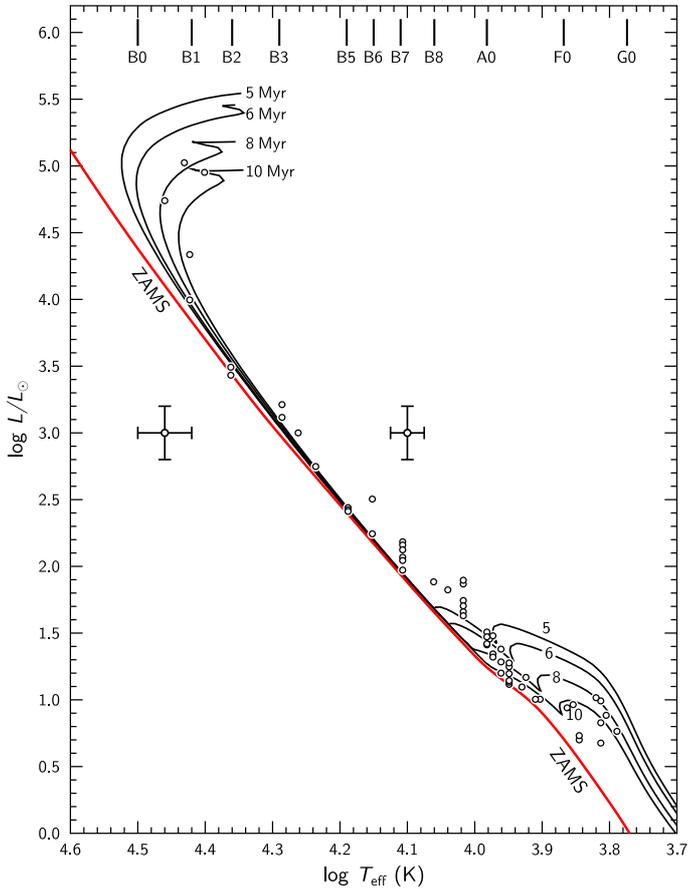}}
\caption{Luminosity vs. effective temperature diagram for 59 probable
members of IC 4996 and the Padova isochrones for ages 5,
6, 8, and 10 Myr of solar metallicity. The $2\sigma$ error crosses for
two spectral classes are shown.}
\vskip1mm
\end{figure}

Two-dimensional spectral types for the selected cluster members in Table
4 are taken either from Table 2 (photometric classification) or from
Table 3 and the literature (spectroscopic classification).  The
agreement between spectral classes from different sources is
satisfactory, the differences are mostly within one decimal subclass.
However, differences in luminosity classes are more significant, and
this required an individual study of each star.  We suspect that in some
cases the classification differences might be caused by the presence of
close optical neighbor stars, in other cases, the stars may be physical
binaries.  It may be useful to recall that the majority of O- and B-type
stars in clusters are found to be binaries; see the review by
\citet{Duchene2013}.  Some B-stars can be misclassified in luminosity
because of their fast axial rotation and resulting deformation.
For the stars of spectral class G, the classification is uncertain
because they are in the pre-main-sequence stage (probably post-T Tauri
stars) and can possess a somewhat peculiar spectral energy distribution.

The mean parallax of the cluster, 0.522\,$\pm$\,0.031 mas, was obtained
by applying the maximum likelihood procedure (Eq.  (1) of CG18) to 30
cluster members of spectral classes B0--B9 selected from Table 4. The
zero-point correction of 0.029 \citep{Lindegren2018} to the {\it Gaia}
DR2 parallaxes was added.  The bright star HD 193007 with the largest
parallax uncertainty was excluded.  The corresponding distance of the
cluster is $1915^{-108}_{+122}$ pc.  By applying the same equation for
71 cluster members (HD 193007 excluded), we obtain almost the same
parallax value, 0.523\,$\pm$\,0.031 mas.  The corresponding distance
modulus of the cluster is $V$--$M_V$ = 11.41 mag.

The histogram for the initially selected 72 cluster members with the
distances from \citet{Bailer2018} is shown as red columns in Figure 3.
The Gaussian curve gives a distance of 1917 pc with an uncertainty
$\sigma$ = 123 pc, which means that the result is very similar to that
obtained from the mean parallax of B-type stars.

Figure 4 shows the physical diagram log\,$L/L_{\odot}$ vs. log\,$T_{\rm
eff}$ for the cluster members of spectral classes B-A-F from Table 4.
The isochrones for ages of 5, 6, 8, and 10 Myr are taken from the Padova
database of stellar evolutionary tracks and isochrones for solar
metallicity \citep{Girardi2002, Bressan2012}.  The ZAMS line was drawn
through the unevolved portions of the isochrones corresponding to 1,
50, and 100 Myr.  Luminosities of stars in solar units were calculated
with the equation
\begin{eqnarray}
\log L/L_{\odot}& = & 0.4 (M_{{\rm bol},\odot} - M_{{\rm bol},\star})=\\
                & & 0.4\,(4.72 - V_0 + DM - BC),\nonumber
\end{eqnarray}
where $V_0 = V - A_V$ is the intrinsic magnitude of the star, $M_{\rm
bol, \odot} = 4.72$ is the absolute bolometric magnitude of the Sun,
$M_{\rm bol, \star} = M_V + BC$ is the absolute bolometric magnitude of
the star, $BC$ is its bolometric correction, and $DM$ is the true
distance modulus of the cluster (11.41 mag).  The effective temperatures
and bolometric corrections of stars were taken according to their
spectral types from Appendix 3 of \citet{Straizys1992}. This $T_{\rm
eff}$ scale is similar to the scales given by \citet{Flower1996},
\citet{Bessel1998}, and \citet{Torres2010}. For G-type pre-main-sequence
members, no values of the physical parameters are given because their
spectral classes are of low accuracy.

In Figure 4, the four most massive members of IC 4996 (HD 193007,
B0\,II; HD 193076, B0.5\,II, BD+37 3862, B0.5\,IV, and BD+37 3859,
B1\,III = No. 573 in Table 2) fit the 8--10 Myr isochrones well.  A few
B3--B9 stars lie slightly above the ZAMS; their luminosity classes are
between V and IV, and some of them are suspected binaries.  However, in
most cases, their deviations from isochrones are within the 2$\sigma$
error crosses, as shown in Figure 4. The isochrones show that the
deviation of cluster stars from the ZAMS starts at spectral class B9.

\section{Relation of IC 4996 to the Cyg OB1 and Cyg OB3 associations}

In most earlier investigations and reviews, the cluster IC 4996 has been
considered as a member of the Cyg OB1 association.  Now we have the
possibility to verify the proposed relationship of the cluster and the
association using astrometric data from the {\it Gaia} DR2 catalogs.
With this aim, we applied the distance histogram to individual stars of
Cyg OB1 taken from \citet{Bailer2018}.  The plot of the distance
histogram for 68 stars of the Cyg OB1 association from
\citet{Humphreys1984} with the reliable parallaxes shows that they are
scattered between 0.9 and 3.3 kpc, which means that many stars of the
suggested members are probably foreground and background stars.  Only 44
stars are concentrated in the distance range between 1.4 and 1.9 kpc,
with a sharp maximum at 1.6--1.7 kpc.  In the VPD of proper motions,
most stars of the Cyg OB1 association are concentrated in the area
containing members of the clusters M29, Berkeley 86, Berkeley 87, and IC
4996.  Nine stars of the histogram peak between 1.4 and 1.9 kpc in the
VPD deviate from the area that contains stars of the clusters listed
above.  Proper motions and radial velocities of stars in associations
usually show much larger dispersion than in open clusters.  We therefore
did not reject these nine stars from the list of the accepted
association members, which is provided in Table 5, and their
histogram is plotted with the blue columns in Figure 3.  The maximum of
the Gaussian curve gives for them $d$ = 1682 pc with $\sigma$ = 116 pc.
At this distance, the diameter of Cyg OB1 is $\sim$\,100 pc.  Thus, the
cluster IC 4996 is too distant to be considered a member of the Cyg OB1
complex.  Its radial distance from the center of Cyg OB1 is larger by a
factor of 4.6 than the radius of the association.

Nine stars of the \citet{Humphreys1984} list of the supposed Cyg OB1
members are concentrated in the distance range 1.95--2.14 kpc, that is,
they are more distant than most of the other stars.  In the sky plane
these stars are located southwest from IC 4996, close to the Cyg OB3
association border.  According to our analysis, most of the Cyg OB3
members in the {\it Gaia} distance histogram show a peak at 1.8--2.0
kpc, which almost coincides with the distance of IC 4996.  Thus, these
nine stars from the Cyg OB1 list, together with the cluster IC 4996, may
be members of Cyg OB3, not of Cyg OB1.  This agrees with the results of
a recent analysis of the kinematic structure of Cyg OB1 by
\citet{Costado2017} based on radial velocity data.  The authors found
that the cluster IC 4996, together with the surrounding stars of the Cyg
OB1 association, form a separate space system at larger distance than
the stars located in the northwest part of Cyg OB1 in the vicinity of
the cluster M29 and close to the association Cyg OB9.

\section{Interstellar extinction in the cluster area}

The plot of the extinction $A_V$ versus {\it Gaia} distance up to $d$ =
4.0 kpc is shown in Figure 5. We plotted only the stars brighter
than $V$ = 16 mag because for the fainter stars the extinction and
distance errors are much larger.  As described in Section 3, distances
to the stars were taken from \citet{Bailer2018}, where they were
determined from the {\it Gaia} DR2 parallaxes.  For most of the stars,
the uncertainties in the negative direction are 50--100 pc at $d$ = 2
kpc and 75--150 pc at $d$ = 3 kpc.  In the positive direction, the
uncertainties are larger by a factor of $\sim$\,1.15.

At a distance of 700--800 pc, the extinction steeply increases to a
value of $\sim$\,1.6 mag.  Most probably, this jump in extinction is
related to the cloud system of the Great Cygnus Rift.  A similar jump
has been found in our earlier papers \citep{Straizys2014, Straizys2015}
in the direction of the nearby cluster M29. The next increase in
extinction is observed at $d$\,$>$\,1.7 kpc.  This dust layer is
probably related to the nearby associations Cyg OB1 and Cyg OB3.  Figure
5 shows the IC 4996 members of spectral classes B and A as red
dots.  Cooler members of the cluster are not plotted since they are
still in the pre-main-sequence stage of evolution, their spectral energy
distributions can be somewhat peculiar, and the extinction values based
on the intrinsic colors can be uncertain.

It is evident that the cluster members cover a broad range of
extinction values, from 1.3 to 2.4 mag.  Most probably, this extinction
variability appears in the foreground of the cluster because there are
no signs that the cluster contains a significant amount of interstellar
dust, which probably was evaporated and moved away after the
formation of high-mass cluster members.  The extinction values are
highest (between 2.0 and 2.4 mag) in the western and southern parts of
the investigated area, and the lowest extinction ($\sim$\,1.3 mag) is
seen in the eastern direction from the cluster center.

In the cluster background, the maximum extinction values are cut by the
limiting magnitudes of the stars.   In Figure 5, the limiting
magnitude $V$ = 16.0 curves for spectral types A0 V and A5 V are shown.
To verify the extinction run at larger distances, the stars to fainter
limiting magnitudes should be plotted.  Our plot of $A_V$ versus
distance for the stars down to $V$ = 18 mag up to $d$ = 7 kpc (not shown
here) shows extinction between 1.5 and 3.0 mag, only a few stars are
seen up to 4 mag.  However, they are of little use for estimating the
mean extinction at different distances because the accuracy of
photometric and {\it Gaia} distances at 5--7 kpc is relatively low.


\begin{figure}
\resizebox{\hsize}{!}{\includegraphics{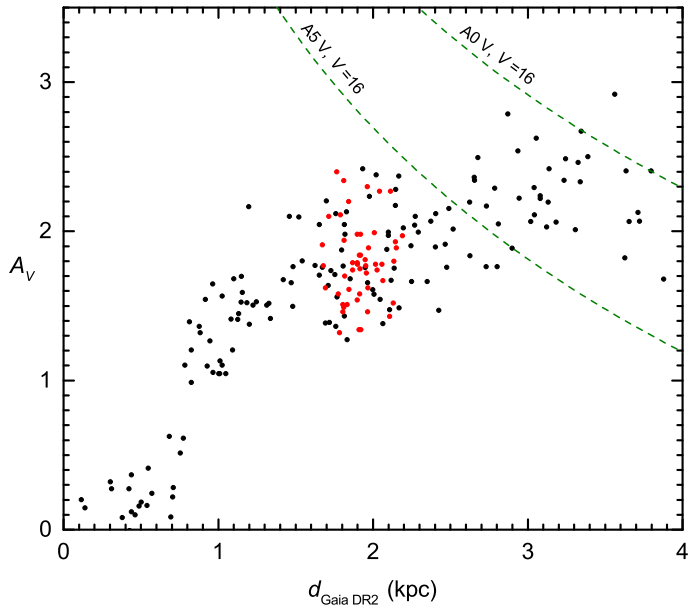}}
\vskip1mm
\caption{Dependence of the extinction on distance for the investigated
area up to $d$ = 4.0 kpc.  The red dots show members of IC 4996 of
spectral classes B and A in the 13$\arcmin$\,$\times$\,13$\arcmin$ area.
The green broken curves show the effect of limiting magnitudes for A0\,V
and A5\,V stars at $V$ = 16.0 mag.}
\vskip1mm
\end{figure}

\section{Conclusions}

We have presented photometry in the Vilnius seven-color system for 1337
stars down to $V$ = 18.8 mag in the area of the open cluster IC 4996.
For 710 stars with the most reliable two-dimensional
classification, the interstellar extinctions were calculated. We
identified 72 possible cluster members by applying their proper motions
from the {\it Gaia} DR2 catalog and the distances based on the {\it
Gaia} parallaxes from \citet{Bailer2018}.  For the brightest stars,
spectroscopic MK types were also obtained from the Asiago spectra or are
available in the literature.  We confirm the cluster sequence from early
B- to G-type stars, where the stars of spectral classes A, F, and G are
still in the pre-main-sequence stage of evolution.

New cluster parameters were derived. The mean parallax of the
cluster (0.522 mas) was obtained by applying the maximum likelihood
procedure to the {\it Gaia} DR2 parallaxes of its 30 B-type members and
adding a systematic error of 0.029 mas.  The resulting parallax
corresponds to a distance of 1915 pc (with $\sigma$\,$\approx$\,110 pc)
and to a true distance modulus of 11.41 mag.  Our distance is similar to
the distance obtained by \citet{Cantat2018} for a smaller area and for a
partly different set of the cluster members.  IC 4996 is probably not a
member of the Cyg OB1 association, which according to the {\it Gaia} DR2
parallaxes, is located at 1682 pc.  More likely, IC 4996 has a common
origin with the association Cyg OB3, which we find to be located at
1.8--2.0 kpc, that is, very close to the cluster.  All this means that
in the post-Gaia era, boundaries of the associations should be
revised taking into account the radial distances of O--B stars of higher
accuracy.  The age of the cluster IC 4996, 8--10 Myr, is estimated from
its physical HR diagram log\,$L/L_{\odot}$ versus log\,$T_{\rm eff}$ and
the Padova isochrones for the solar metallicity.

The interstellar extinction parameter $A_V$ in the direction of cluster
members covers a wide range of values, from 1.3 to 2.4 mag.  The plot of
$A_V$ versus $d$ in the area of IC 4996 shows a steep increase in
extinction of up to 1.6 mag at 700--800 pc, which is probably related to
dust clouds at the edge of the Great Cygnus Rift.  The increase in
extinction of up to 2.4 mag at $d$\,$\geq$\,1.7 kpc may be related to
the Cyg OB1 and Cyg OB3 associations.

This study is an example of how a two-dimensional classification of
stars based on multicolor photometry, combined with the {\it Gaia}
astrometric data, can be applied for obtaining physical parameters of
open clusters.  For the further improvement of the accuracy of cluster
parameters, a better recognition of unresolved binaries and peculiar
stars is desirable.  For B-type stars, it is important to take the
effects of axial rotation into account.  The measurements and analysis
of radial velocities and profiles of spectral lines for early-type stars
would be helpful in detecting binarity and rotation.

{\bf Acknowledgments}.  This work has made use of data from the European
Space Agency (ESA) mission Gaia (https://www.cosmos.esa.int/gaia),
processed by the Gaia Data Processing and Analysis Consortium (DPAC,
https://www.cosmos.esa.int/web/gaia/dpac/consortium).  Funding for the
DPAC has been provided by national institutions, in particular the
institutions participating in the Gaia Multilateral Agreement.  The use
of the Simbad (CDS), WEBDA (Masaryk University) and SkyView (NASA)
databases is also acknowledged.  Preliminary results of this
investigation were presented at the AAS Meeting No. 231
\citep{Boyle2018}.  We are grateful to Carme Jordi and Tristan
Cantat-Gaudin for submitting their results before publication and for
important comments.  The project is partly supported by the Research
Council of Lithuania, grant No.  S-MIP-17-74.

\bibliographystyle{aa}
\bibliography{ref4996}

\end{document}